\documentstyle[prd,eqsecnum,aps,preprint]{revtex}
\tighten
\begin{document}
\draft
\preprint{
EPHOU-98-008}
\title{Note on the field theory of neutrino mixing}
\author{Kanji Fujii$^{*1}$, Chikage Habe$^{*1}$ and Tetsuo Yabuki$^{*2}$}
\address{$^{*1}$ Department of Physics, Faculty of Science,\\ 
\hspace*{8mm}Hokkaido University, Sapporo 060-0810, Japan\\
$^{*2}$ Rakunou Gakuen University, Ebetsu 069-0836, Japan}
\maketitle
\begin{abstract}
The problem of the diagonalization of the flavor-neutrino propagator matrix is investigated in the theory with flavor-mixing mass terms in Lagrangian.
For this purpose we examine one-pole structures of flavor-neutrino propagators, leading to physical neutrino masses, and discuss the relation of the propagator diagonalizaion to the diagonalizaion of the mass matrix in Lagrangian. 
In connection with the paper by Blasone et al., it is pointed out that there is no compelling reason for fixing the mass parameters, althought this fixing is necessary in order to construct the flavor Hilbert space.
\end{abstract}

\section{Introduction}\label{sec1}
Since Pontecorvo\cite{pontecorvo} pointed out the possibility of the neutrino oscillation and, in addition, the solar neutrino problem was proposed\cite{davis}, the oscillation has been much investigated experimentally 
and theoretically.
Indications in favor of the neutrino oscillation from various kinds of experiments have been reported\cite{bilenky}.

The main aim of the present note is to investigate the field theory of neutrino mixing and to give a remark on the way how to define the physical neutrino masses on the basis of Green-function approach, which has not been noticed in current literatures. 
In Sec.\ref{sec2}, we investigate this problem in the two-flavor case as an simple illustration, and Sec.\ref{sec3} the three-flavor case is to be examined.

In Sec.\ref{sec4}, we give some remarks on the field theory of neutrino mixing and the construction of the Fock space of definite flavors.
For the later convenience, here we summerize the problems included in such a field-theoretical approach.
Blasone and Vitiello\cite{BV} have considered the field theory of neutrino mixing. 
Their consideration is based on the unitary inequivalence of the Fock space for definite flavor states to that for definite mass states. 
Although the investigated theme is very interesting\cite{fujii}, there are some problematic points in ref.\cite{BV}. 
The first point is that the expansion of the fermion field in terms of plane-wave eigenfunctions of the third component of the spin operator is employed. 
Such a way of the expansion causes an unnecessary and nonessential complications in presentation of the theory. 
When the plane-wave eigenfunctions of the helicity are employed from the outset, as finally done also in ref.\cite{BV}, the description of the theory will be much simplified and becomes clearer.

Whereas the first point mentioned above is of technical character, the second point is more fundamental. 
The second point is related to the problem how to define the annihilation and creation operators for definite flavors and definite masses. 
The relations from (2.23a) to (2.23d) given in ref.\cite{BV} are written as
\begin{eqnarray}
u_{\sigma}(k,r)\tilde\alpha_{\sigma}(k,r)=G^{-1}(\theta)u_{j}(k,r)\alpha_{j}(k,r)G(\theta)\label{1.1}\\
v_{\sigma}^{*}(k,r)\tilde\beta_{\sigma}(k,r)=G^{-1}(\theta)v_{j}^{*}(k,r)\beta_{j}(k,r)G(\theta)\label{1.2}
\end{eqnarray}
where (in the case of the two flavors) $\{\sigma, j\}=\{e,1\}$ and $\{\mu,2\}$; $u_{b}(k,r)$, $b=\sigma \mbox{ or } j$, is the plane \-wave eigenfunction with the mass $m_{b}$ and the spin component $r$; $G(\theta)$ is given by
\begin{eqnarray}
G(\theta)=exp\left[\theta\!\int\! d^{3}x (\nu_{1}^{\dag}(x)\nu_{2}(x)-\nu_{2}^{\dag}(x)\nu_{1}(x))\right].\label{1.3}
\end{eqnarray}
Under the excuse of convenience, $u_{\sigma}(k,r)$ and $v_{\sigma}(k,r)$ have been chosen to be the eigenfunctions with the masses $m_{1}$ and $m_{2}$ for $\sigma =e$ and $\mu$, respectively;
that is, the authors of ref.\cite{BV} defined the operators $\alpha^{BV}_{\sigma}$ and $\beta^{BV}_{\sigma}$ through $u_{j}\alpha^{BV}_{\sigma}\equiv u_{\sigma}\tilde\alpha_{\sigma}$ and $v^{*}_{j}\beta^{BV}_{\sigma}\equiv v^{*}_{\sigma}\tilde\beta_{\sigma}$.
Thus, from (\ref{1.1}) and (\ref{1.2}), they obtained 
\begin{eqnarray}
\left(\begin{array}{c}\alpha^{BV}_{\sigma}(k,r)\\ \beta^{BV}_{\sigma}(k,r)\end{array}\right)=G^{-1}(\theta)
\left(\begin{array}{c}\alpha_{j}(k,r)\\ \beta_{j}(k,r)\end{array}\right)G(\theta).\label{1.4}
\end{eqnarray}

Although the relation (\ref{1.1}) and (\ref{1.2}) employed for deriving (\ref{1.4}) are not understandable, this last relation can be understood as follows. 
Any Heisenberg field $\nu_{b}(x)$ can be expanded\cite{umezawa} in terms of the helicity-momentum eigenfunctions\cite{cornaldesi} as
\begin{eqnarray}
\nu_{b}(x)=\frac{1}{\sqrt{V}}\sum_{\vec k,r}e^{i\vec k\vec x}\{ u_{b}(k,r)\alpha_{b}(k,r;t)+
v_{b}(-k,r)\beta^{\dagger}_{b}(-k,r;t)\},\mbox{\hspace*{5mm}} t=x^{0}.
\label{1.5}\end{eqnarray}
Here, $u_{b}(k,r)$ and $v_{b}(k,r)$ satisfy
\begin{eqnarray}
(i\not k +m_{b})u_{b}(k,r)=0,\mbox{\hspace*{5mm}} (i\not k -m_{b})v_{b}(k,r)=0,\mbox{\hspace*{5mm}}k_{0}=\sqrt{\vec k^{2}+m^{2}_{b}},
\label{1.6}\end{eqnarray}
where $\not k=\gamma^{\alpha}k_{\alpha}=\vec\gamma\vec k+\gamma^{4}ik_{0}$;
$v_{b}(-k,r)$ and $\beta^{\dagger}_{b}(-k,r;t)$ are the quantities with the 3-momentum $-\vec k$;
the helicity eigenfunctions are used for technical simplicity in the following, and their concrete forms are given in Appendix A.
The expansion coefficient operators in (\ref{1.5}) satisfy the canonical commutation relations for the equal time, which are derived from the equal-time commutation relations, $\{\nu_{b}(x),\nu^{\dagger}_{b'}(y)\}=\delta(\vec x-\vec y)\}\delta_{bb'}$ and $\mbox{others }=0$.
From the relation
\begin{eqnarray}
\nu_{\sigma}(x)=G^{-1}(\theta;t)\nu_{j}(x)G(\theta;t),\mbox{\hspace*{5mm}}t=x^{0},
\label{1.7}\end{eqnarray}
we obtain
\begin{eqnarray}
\{u_{\sigma}(k,r)\alpha_{\sigma}(k,r;t)&+&
v_{\sigma}(-k,r)\beta^{\dagger}_{\sigma}(-k,r;t)\}\nonumber\\
&=&G^{-1}(\theta;t)\{u_{j}(k,r)\alpha_{j}(k,r;t)+
v_{j}(-k,r)\beta^{\dagger}_{j}(-k,r;t)\}G(\theta;t),\label{1.8}
\end{eqnarray}
where $\{u_{\sigma},v_{\sigma}\}$ and $\{u_{j},v_{j}\}$ are the plane-wave eigenfunctions with masses $m_{\sigma}$ and $m_{j}$, respectively.
(\ref{1.8}) is the relation which can be utilized instead of (\ref{1.1}) and (\ref{1.2}), and leads to the general linear transformation between $\{\alpha_{\sigma}(k,r), \beta^{\dagger}_{\sigma}(-k,r),\sigma=e,\mu\}$ and $\{\alpha_{j}(k,r), \beta^{\dagger}_{j}(-k,r),j=1,2\}$ as
\begin{eqnarray}
\left(\begin{array}{c}\alpha_{e}(k,r)\\\alpha_{\mu}(k,r)\\\beta^{\dagger}_{e}(-k,r)\\\beta^{\dagger}_{\mu}(-k,r)\end{array}\right)={\cal G}(\theta)
\left(\begin{array}{c}\alpha_{1}(k,r)\\\alpha_{2}(k,r)\\\beta^{\dagger}_{1}(-k,r)\\\beta^{\dagger}_{2}(-k,r)\end{array}\right)
, \mbox{\hspace*{5mm}}{\cal G}(\theta)^{\dagger}={\cal G}(\theta)^{-1},\label{1.9}
\end{eqnarray}
The concrete form of ${\cal G}(\theta)$ is easily obtained as given in Appendix A by utilizing the explicit forms of $u_{\sigma}, v_{\sigma}, u_{j}$ and $v_{j}$\cite{cornaldesi}. 
If we take $m_{e}=m_{1}$ and $m_{\mu}=m_{2}$ as a special case of (\ref{1.9})(or (\ref{1.8})), we obtain the relation (\ref{1.4}), which plays the basic role in ref.\cite{BV}.
Thus it is necessary for us to make clear a logical basis of the above choice of $m_{e}$ and $m_{\mu}$, on which we examine in Sec.\ref{sec4}.

Giunti et al.\cite{giunti} assert that it is impossible to define generally the creation and annihilation operators with a definite flavor, and that the construction of the Fock space of 'weak'(or flavor) states is approximately allowed only in the extremely relativistic case.
Along a different context, we will obtain the same conclusion, although the second assertion mentioned above is evident, since in (\ref{1.9}) all dependences of ${\cal G}(\theta,k)$ on mass diferences among $m_{j}$'s and $m_{\sigma}$'s become negligible and we obtain
\begin{eqnarray}
\left(\begin{array}{c}\alpha_{e}(k,r;t)\\ \alpha_{\mu}(k,r;t)\end{array}\right)=
\left(\begin{array}{cc}cos\theta&sin\theta\\-sin\theta&cos\theta\end{array}\right)
\left(\begin{array}{c}\alpha_{1}(k,r;t)\\ \alpha_{2}(k,r;t)\end{array}\right)\label{1.10}
\end{eqnarray}
as well as the same relation between $(\beta_{e}(k,r), \beta_{\mu}(k,r))$and $(\beta_{1}(k,r), \beta_{2}(k,r))$. 
In Appendix B we add a related remark.

\section{diagonalization of the neutrino propagator -----the case of 2-flavors -----}\label{sec2}
We examine the diagonalization of the neutrino propagator according to the procedure proposed by Kaneko, Ohnuki and Watanabe\cite{kaneko}, which had been developed many years ago as the field theory of particle mixture interaction. 
In this section we consider the two-flavor case as an illustration.
(In this paper we confine ourselves to the case of Dirac neutrino.)

\subsection{Starting Lagrangian}
Let us consider the following Lagrangian density with a mutual transition between two neutrino fields specified by the flavor degrees of freedom $\sigma=e$ and $\mu$;
\begin{eqnarray}
{\cal L}=-\left(\begin{array}{cc}\bar{\nu}_{e}(x)&\bar{\nu}_{\mu}(x)\end{array}\right)
(\not{\partial}+M)
\left(\begin{array}{c}\nu_{e}(x)\\\nu_{\mu}(x)\end{array}\right) 
+ {\cal L}_{int},
\label{2.1}\end{eqnarray}
where
\begin{eqnarray}
M=
\left(\begin{array}{cc}m_{ee}&m_{e\mu}\\m_{\mu e}&m_{\mu\mu}\end{array}
\right);\mbox{\hspace*{5mm}}\not{\partial}:=\gamma^{\rho}\partial_{\rho}=\vec{\gamma}\vec{\nabla}+\gamma^{4}\frac{1}{i}\frac{\partial}{\partial x^{0}}, \mbox{\hspace*{5mm}}(\gamma^{\rho})^{\dagger}=\gamma^{\rho}.\label{2.2}
\end{eqnarray}
Due to $M^{\dagger}=M$, required from the hermiticity of ${\cal L}(x)$, $m_{ee}$ and $m_{\mu\mu}$ are real and $m_{e\mu}^{*}$ is equal to $m_{\mu e}$. 
${\cal L}_{int}$ in (\ref{2.1}) is assumed to have no bilinear terms and no derivatives of the neutrino field operators; then the Hamiltonian is
\begin{eqnarray}
{\cal H}(x)={\cal H}^{0}_{e\mu}(x)-{\cal L}_{int}(x)\label{2.3}
\end{eqnarray}
with
\begin{eqnarray}
{\cal H}^{0}_{e\mu}(x)=
\left(\begin{array}{cc}\bar{\nu}_{e}(x)&\bar{\nu}_{\mu}(x)\end{array}\right)\label{2.4}
(\vec\gamma \vec\nabla+M)
\left(\begin{array}{c}\nu_{e}(x)\\\nu_{\mu}(x)\end{array}\right).
\end{eqnarray}
The eigenvalues of $M$ are
\begin{eqnarray}
m_{\stackrel{\scriptstyle 1}{(2)}}=\frac{1}{2}\left(m_{ee}+m_{\mu\mu}
\begin{array}{c}-\\(+)\end{array}
\sqrt{(m_{\mu\mu}-m_{ee})^{2}+4|m_{e\mu}|^{2}}\right),\label{2.5}
\end{eqnarray}
and ${\cal H}^{0}_{e\mu}(x)$ is expressed in the diagonalized form as
\begin{eqnarray}
{\cal H}^{0}_{e\mu}(x)=
\sum\bar{\nu}_{j}(x)(\vec\gamma \vec\nabla+m_{j})\nu_{j}(x).\label{2.6}
\end{eqnarray}
For simplicity, we take $m_{e\mu}=m_{\mu e}$, derived from CP-invariance; then we can take
\begin{eqnarray}
\left(\begin{array}{c}\nu_{e}(k,r)\\ \nu_{\mu}(k,r)\end{array}\right)=
\left(\begin{array}{cc}cos\theta&sin\theta\\-sin\theta&cos\theta\end{array}\right)
\left(\begin{array}{c}\nu_{1}(k,r)\\ \nu_{2}(k,r)\end{array}\right)\label{2.7}
\end{eqnarray}
with
\begin{eqnarray}
tan\theta=\frac{1}{2m_{e\mu}}\left(-(m_{\mu\mu}-m_{ee})+\sqrt{(m_{\mu\mu}-m_{ee})^{2}+4m_{e\mu}^{2}}\right).\label{2.8}
\end{eqnarray}
We take $m_{\mu\mu}\geq m_{ee}\geq 0$ with no loss of generality; then
$m_{2}\geq |m_{1}|$ and 
\begin{eqnarray}
m_{1}\geq 0 \mbox{\hspace*{3mm} for \hspace*{3mm}} \sqrt{m_{ee}m_{\mu\mu}}\geq |m_{e\mu}|,\mbox{\hspace*{5mm}}m_{1}< 0 \mbox{\hspace*{3mm} for \hspace*{3mm}} \sqrt{m_{ee}m_{\mu\mu}}< |m_{e\mu}|. \label{2.9}
\end{eqnarray}

In the following calculations, it will be useful for us to employ the relations
\begin{eqnarray}
m_{ee}&=&m_{1}(cos\theta )^{2}+m_{2}(sin\theta )^{2}, \mbox{\hspace*{5mm}} 
m_{\mu\mu}=m_{1}(sin\theta )^{2}+m_{2}(cos\theta )^{2},\label{2.10}\\
m_{e\mu}&=&sin\theta cos\theta (-m_{1}+m_{2}), \mbox{\hspace*{5mm}} tan(2\theta)=\frac{2m_{e\mu}}{m_{\mu\mu}-m_{ee}};\label{2.11}
\end{eqnarray}
further we have
\begin{eqnarray}
m_{1}m_{2}&=&m_{ee}m_{\mu\mu}-m_{e\mu}^{2},\label{2.12}\\ 
m_{ee}-m_{1}&=&-m_{\mu\mu}+m_{2}=m_{e\mu}tan\theta,\label{2.13}\\
m_{ee}-m_{2}&=&-m_{\mu\mu}+m_{1}=-m_{e\mu}cot\theta.\label{2.14}
\end{eqnarray}
\subsection{Poles of the propagator matrix}\label{sec2.2}
With the aim of examining the propagation character of $\nu_{\sigma}$-field, we consider the propagator
\begin{eqnarray}
S'_{\sigma\rho}(x-y):=<0|T(\nu_{\sigma}(x)\bar\nu_{\rho}(y))|0>,\label{2.15}
\end{eqnarray}
where $\nu_{\sigma}(x)$ is the flavor neutrino field appearing in the Lagrangian (\ref{2.1}) and is called the unrenormalized Heisenberg operator in accordance with the Lehman's terminology\cite{lehmann}.
It is necessary for us to define the vacuum $|0>$.
Here we assume that, corresponding to a given Hamiltonian, the vacuum with the lowest energy exists. 
The Fourier transform of the propagator (\ref{2.15}), $S'_{\sigma\rho}(\not {k})=\int d^{4}xexp(-ikx)S'_{\sigma\rho}(x)$, satisfies
\begin{eqnarray}
S'_{\sigma\rho}=\delta_{\sigma\rho}S_{\rho}+\sum_{\lambda}S'_{\sigma\lambda}\Pi_{\lambda\rho}S_{\rho},\label{2.16}
\end{eqnarray}
where $S_{\rho}(\not{k}):=(-\not{k}+im_{\rho\rho})^{-1}$ is the free propagator of the $\nu_{\sigma}$-field.
When we define the matrix $[f_{\sigma\rho}(\not{k})]$ to be 
\begin{eqnarray}
S'_{\sigma\rho}(\not{k})=[f(\not{k})^{-1}]_{\sigma\rho}\label{2.17}
\end{eqnarray}
we obtain 
\begin{eqnarray}
f_{\sigma\rho}(\not{k})=\delta_{\sigma\rho}S_{\rho}(\not{k})^{-1}-\Pi_{\sigma\rho}(\not{k}).\label{2.18}
\end{eqnarray}
$[S'_{\sigma\rho}(\not {k})]$ has two poles, determined by 
\begin{eqnarray}
det[f_{\sigma\rho}(\not{k})]=0.\label{2.19}
\end{eqnarray}

Let us examine the pole structure of $S'_{\sigma\rho}$ under the approximation for the proper self-energy part $\Pi_{\sigma\rho}$ by neglecting ${\cal L}_{int}(x)$ in (\ref{2.3}) and by taking into account only the contribution from e $\rightarrow\!\!\!\!-\!\!-\!\!\!\!\!\!\stackrel{m_{e\mu}}{\times}\!\!\!\!\!\!-\!\!-\!\!\!\!\rightarrow \mu$ in the lowest order; thus, we have
\begin{eqnarray}
[f_{\sigma\rho}(\not{k})]=
\left(\begin{array}{cc}
-{\not k}+im_{ee}&im_{e\mu}\\
im_{e\mu}&-{\not k}+im_{\mu\mu}
\end{array}\right).\label{2.20}
\end{eqnarray}
As easily seen, $[S'_{\sigma\rho}(\not{k})]$ has two poles at
\begin{eqnarray}
\not k =im_{j} \mbox{\hspace*{5mm} with } m_{j}(j=1,2) \mbox{ given by (\ref{2.5})}.\label{2.21}
\end{eqnarray}
Therefore, the physical one-particle masses given as poles of $S'_{\sigma\rho}(\not{k})$ are seen to coincide with the eigenvalues of the mass matrix $M$.

It should be noted that there is an arbitrariness in separating ${\cal H}(x)$ into the "free" and "interaction" parts.
So it is worthy to give a remark on this point.
We rewrite the Hamiltonian (\ref{2.4}) as
\begin{eqnarray}
{\cal H}^{0}_{e\mu}(x)=
\left(\begin{array}{cc}\bar{\nu}_{e}(x)&\bar{\nu}_{\mu}(x)\end{array}\right)
(\not{\partial}+\left(
\begin{array}{cc}
m^{0}_{ee}&0\\
0&m^{0}_{\mu\mu}
\end{array}
\right))
\left(\begin{array}{c}
\nu_{e}(x)\\
\nu_{\mu}(x)
\end{array}\right) + H_{int}(x), \label{2.22}\\
H_{int}(x)=
\left(
\begin{array}{cc}
\Delta_{ee}&m_{e\mu}\\
m_{e\mu}&\Delta_{\mu\mu}
\end{array}
\right), \mbox{\hspace*{5mm}}\Delta_{\sigma\sigma}:=m_{\sigma\sigma}-m^{0}_{\sigma\sigma};\label{2.23}
\end{eqnarray}
Then, instead of $S_{\sigma}$ and $\Pi_{\rho\sigma}$ employed above, we use
\begin{eqnarray}
S^{0}_{\rho}(\not{k}):=(-\not{k}+im^{0}_{\rho\rho})^{-1}, \mbox{\hspace*{5mm}}
[\Pi^{0}_{\rho\sigma}]:=\left(
\begin{array}{cc}
-i\Delta_{ee}&-im_{e\mu}\\
-im_{e\mu}&-i\Delta_{\mu\mu}
\end{array}
\right).\label{2.24}
\end{eqnarray} 
Dropping contributions from ${\cal L}_{int}(x)$ to the proper self-energy part and taking account of the contribution from $H_{int}(x)$, we obtain
\begin{eqnarray}
\delta_{\sigma\rho}S^{0}_{\rho}-\Pi^{0}_{\sigma\rho}=\delta_{\sigma\rho}S_{\rho}-\Pi_{\sigma\rho}=f_{\sigma\rho},\label{2.25}
\end{eqnarray}
which shows the arbitrariness in defining $S_{\rho}(\not k)$ disappears in the physical one-particle masses.
\subsection{Diagonalization of the pole part in the propagator}
We examine diagonalization of the pole part in the neutrino propagator $S'_{\sigma\rho}(\not k)$.
Writing the cofactor corresponding to $f_{\sigma\rho}$ as $F_{\sigma\rho}$, we have
\begin{eqnarray}
[S'_{\sigma\rho}]=[f_{\sigma\rho}]^{-1}=\frac{1}{det[f]}[F_{\sigma\rho}]^{T}.\label{2.26}
\end{eqnarray}
We define $f_{\sigma\rho}^{(j)}$ and $F_{\sigma\rho}^{(j)}$ to be the values of $f_{\sigma\rho}$ and $F_{\sigma\rho}$ at the pole $m_{j}$ of $[S'_{\sigma\rho}]$, respectively.
We have
\begin{eqnarray}
[f_{\sigma\rho}^{(j)}][F_{\sigma\rho}^{(j)}]^{T}&=&
\left(
\begin{array}{cc}
i(-m_{j}+m_{ee})&im_{e\mu}\\
im_{e\mu}&i(-m_{j}+m_{\mu\mu})
\end{array}
\right)
\left(
\begin{array}{cc}
i(-m_{j}+m_{\mu\mu})&-im_{e\mu}\\
-im_{e\mu}&i(-m_{j}+m_{ee})
\end{array}
\right)\nonumber \\
&=&[-(m_{j}-m_{ee})(m_{j}-m_{\mu\mu})+m_{e\mu}^{2}]
\left(
\begin{array}{cc}
1&0\\
0&1
\end{array}
\right)
=det[f^{j}]\left(
\begin{array}{cc}
1&0\\
0&1
\end{array}
\right)
=0\label{2.27}
\end{eqnarray}
due to (\ref{2.13}) and (\ref{2.14}) (or in accordance with (\ref{2.19})).

Next we define
\begin{eqnarray}
(\rho^{(j)})^{-1}:=\frac{d det[f_{\sigma\rho}(\not k)]}{d(-\not k)}|_{\not k =im_{j}}.\label{2.28}
\end{eqnarray}
From (\ref{2.20}) we obtain
\begin{eqnarray}
\mbox{r.h.s. of (\ref{2.28})}=[-2\not k +i(m_{ee}+m_{\mu\mu})]_{\not k =im_{j}}=i(m_{1}-m_{2})(-1)^{j},\label{2.29}
\end{eqnarray}
leading to
\begin{eqnarray}
\left( \begin{array}{c} \rho^{(1)}\\ \rho^{(2)} \end{array} \right)=
\left( \begin{array}{c} \frac{1}{i(m_{2}-m_{1})}\\ \frac{1}{i(m_{1}-m_{2})} \end{array} \right)=
\frac{s_{\theta}c_{\theta}}{im_{e\mu}}
\left( \begin{array}{c} 1\\ -1 \end{array} \right), \mbox{\hspace*{3mm}}s_{\theta}=sin\theta , \mbox{\hspace*{3mm}}c_{\theta}=cos\theta .\label{2.30}
\end{eqnarray}
We introduce a set of new fields $\psi^{r}_{j}(x)$ and $\bar\psi^{r}_{j}(x), j=1,2$, expressed as
\begin{eqnarray}
\psi_{\sigma}(x)=\sum_{j}A_{\sigma j}\psi^{r}_{j}(x), \mbox{\hspace*{5mm}}\bar\psi_{\sigma}(x)=\sum_{j}\bar A_{\sigma j}\bar\psi^{r}_{j}(x),\label{2.31}
\end{eqnarray}
where the coefficients $A_{\sigma j}$'s are so determined that $<0|T(\psi^{r}_{i}(x)\bar\psi^{r}_{j}(y))|0>$ has only one pole term like $\delta_{ij}/(-\not k+im_{j})$; $\bar A_{\sigma j}$ is a complex conjugate to $A_{\sigma j}$.
Thus, from the conditions
\begin{eqnarray}
\rho^{(1)}F^{(1)T}=A\left( \begin{array}{cc} 1&0\\ 0&0 \end{array} \right)A^{\dagger}, \label{2.32}\\
\rho^{(2)}F^{(2)T}=A\left( \begin{array}{cc} 0&0\\ 0&1 \end{array} \right)A^{\dagger}, \label{2.33}
\end{eqnarray}
we obtain
\begin{eqnarray}
\rho^{(j)}F^{(j)}_{\rho\sigma}=\bar A_{\rho j}A_{\sigma j},\label{2.34}
\end{eqnarray}
leading to
\begin{eqnarray}
\bar\rho^{(j)}\bar F^{(j)}_{\rho\sigma}=\bar A_{\sigma j}A_{\rho j}=\rho^{(j)}F^{(j)}_{\sigma\rho}.\label{2.35}
\end{eqnarray}
Thus we have 
\begin{eqnarray}
\bar A_{\rho j}A_{\sigma j}=\frac{\bar A_{\rho j}A_{\mu j}\bar A_{\mu j}A_{\sigma j}}{A_{\mu j}\bar A_{\mu j}}=
\left\{ \begin{array}{l} \frac{\bar\rho^{(j)}\bar F^{(j)}_{\mu\rho}F^{(j)}_{\mu\sigma}}{F^{(j)}_{\mu\mu}}\\\mbox{or}\\ \frac{\rho^{(j)}F^{(j)}_{\rho\mu}\bar F^{(j)}_{\sigma\mu}}{\bar F^{(j)}_{\mu\mu}}  \end{array} \right.
\label{2.36}\end{eqnarray}
A possible solution is given by
\begin{eqnarray}
A_{\sigma j}=\frac{|\rho^{(j)}|}{\sqrt{\rho^{(j)}F^{(j)}_{\mu\mu}}}F^{(j)}_{\mu\sigma}\omega, \mbox{\hspace*{5mm}}|\omega|^{2}=1\label{2.37}
\end{eqnarray}
The concrete form of $[A_{\sigma j}]$ is expressed as
\begin{eqnarray}
[A_{\sigma j}]&=&
\left( \begin{array}{cc} 
\frac{|s_{\theta}c_{\theta}/m_{e\mu}|}{\sqrt{\frac{s_{\theta}c_{\theta}}{m_{e\mu}}(-m_{1}+m_{ee})}}
\left( \begin{array}{c} -im_{e\mu}\\ i(-m_{1}+m_{ee}) \end{array} \right), &
\frac{|s_{\theta}c_{\theta}/m_{e\mu}|}{\sqrt{\frac{s_{\theta}c_{\theta}}{m_{e\mu}}(m_{2}-m_{ee})}}
\left( \begin{array}{c} -im_{e\mu}\\ i(-m_{2}+m_{ee}) \end{array} \right)
\end{array} \right)\omega\label{2.38}\\
&=&\frac{m_{e\mu}}{|m_{e\mu}|}i\omega\left( \begin{array}{cc} -|c_{\theta}|\mbox{\hspace*{5mm}}&-|s_{\theta}|\\ |c_{\theta}|\frac{s_{\theta}}{c_{\theta}}\mbox{\hspace*{5mm}}&-|s_{\theta}|\frac{c_{\theta}}{s_{\theta}} 
\end{array} \right).\label{2.39}
\end{eqnarray}
For $s_{\theta},c_{\theta},m_{e\mu}>0$ and $\omega=i$, we have
\begin{eqnarray}
\left[A_{\sigma j}\right]=\left( \begin{array}{cc} c_{\theta}&s_{\theta}\\ -s_{\theta}&c_{\theta} \end{array} \right).\label{2.40}
\end{eqnarray}

From the construction, we see the propagator
\begin{eqnarray}
S^{r}_{ij}(\not k):&=&\mbox{Fourier transform of}<0|T(\psi^{r}_{i}(x)\bar\psi^{r}_{j}(y))|0>\label{2.41}\\
&=&\sum_{\sigma,\rho}B_{i\sigma}S'_{\sigma\rho}(\not k)\bar B_{j\rho}, \mbox{\hspace*{5mm}}B:=A^{-1}\nonumber
\end{eqnarray}
has an one-pole term in diagonal element; i.e.
\begin{eqnarray}
[S^{r}_{ij}(\not k)]&=&[\sum_{l}\frac{1}{-\not k+im_{l}+\epsilon}(\sum_{\sigma,\rho}B_{i\sigma}\rho^{(l)}F^{(l)}_{\sigma\rho}(\not k)\bar B_{j\rho})+(\mbox{contribution from continuous spectra})]\nonumber \\
&=&\left( \begin{array}{cc} \frac{1}{-\not k+im_{1}+\epsilon}&0\\ 0&\frac{1}{-\not k+im_{2}+\epsilon} \end{array} \right)+(\mbox{contribution from continuous spectra}).\label{2.42}
\end{eqnarray}
When we write $A_{\sigma j}$ as $z^{\frac{1}{2}}_{\sigma j}$, i.e.
\begin{eqnarray}
[z^{\frac{1}{2}}_{\sigma j}]:=[A_{\sigma j}],\label{2.43}
\end{eqnarray}
we can express $S'_{\sigma\rho}(\not k)$ as
\begin{eqnarray}
S'_{\sigma\rho}(\not k)=\sum_{j}\frac{z^{\frac{1}{2}}_{\sigma j}\bar z^{\frac{1}{2}}_{\rho j}}{-\not k+im_{j}+\epsilon}+\int d(\kappa^{2})\frac{\lambda_{\sigma\rho}(\kappa,\not k)}{k^{2}+\kappa^{2}-i\epsilon}.\label{2.44}
\end{eqnarray}

Note that the diagonalization procedure of the propagator $S'_{\sigma\rho}$ described above is somewhat different from that adopted by Kaneko et al.\cite{kaneko}.
The authors of ref.\cite{kaneko} considered the intermediate step by introducing a set of fields $\{\phi_{j}(x), \tilde\phi_{j}(x)\}$ as defined by
\begin{eqnarray}
\psi_{\sigma}(x)=\sum_{j}A_{\sigma j}\phi_{j}(x),\nonumber \\
\bar\psi_{\sigma}(x)=\sum_{j}A_{\sigma j}\tilde\phi_{j}(x),\label{2.45}
\end{eqnarray}
and examined the pole-part diagonalization of
\begin{eqnarray}
\tilde S_{ij}(\not k):=\mbox{Fourier transform of}<0|T(\phi_{i}(x)\tilde\phi_{j}(y))|0>\label{2.46}
\end{eqnarray}
We have shown that such an intermediate procedure is not always necessary in order to obtain (\ref{2.42}) and (\ref{2.44}).
\section{Case of three flavors}\label{sec3}

We examine the diagonalization of the flavor neutrino propagator in the three-flavor case along the same line of thought as given in the preceding section.
\subsection{3-flavor mixing mass-matrix}

The relevant Lagrangian density with mutual transitions among three-flavor neutrinos, $e$,$\mu$ and $\tau$, is written after taking account of the spontaneous symmetry breaking in the Higgs sector as
\begin{eqnarray}
{\cal L}(x)=&-&\left( \begin{array}{ccc} \bar\nu_{eL}(x)&\bar\nu_{\mu L}(x)&\bar\nu_{\tau L}(x) \end{array} \right)(\not\partial + M')\left( \begin{array}{c} \nu'_{eR}(x)\\ \nu'_{\mu R}(x)\\ \nu'_{\tau R}(x) \end{array} \right) \nonumber \\
&-&\left( \begin{array}{ccc} \bar\nu'_{eR}(x)&\bar\nu'_{\mu R}(x)&\bar\nu'_{\tau R}(x) \end{array} \right)(\not\partial + M^{\prime\dagger})\left( \begin{array}{c} \nu_{eL}(x)\\ \nu_{\mu L}(x)\\ \nu_{\tau L}(x) \end{array} \right)+{\cal L}'_{int}(x)\label{3.1}
\end{eqnarray}
where $M'=[m'_{\sigma\rho}]$. (${\cal L}'_{int}$ is assumed to include no bilinear terms and no derivative of the neutrino field.)
We perform unitary transformations
\begin{eqnarray}
\left( \begin{array}{c} \nu_{eL}(x)\\ \nu_{\mu L}(x)\\ \nu_{\tau L}(x) \end{array} \right)=V_{L}\left( \begin{array}{c} \nu_{1L}(x)\\ \nu_{2L}(x)\\ \nu_{3L}(x) \end{array} \right), \mbox{\hspace*{5mm}}
\left( \begin{array}{c} \nu_{eR}(x)\\ \nu_{\mu R}(x)\\ \nu_{\tau R}(x) \end{array} \right)=V_{R}\left( \begin{array}{c} \nu_{1R}(x)\\ \nu_{2R}(x)\\ \nu_{3R}(x) \end{array} \right), \label{3.2a}
\end{eqnarray}
so that the mass matrix is diagonalized;
\begin{eqnarray}
V_{L}^{\dagger}M'V_{R}=\left( \begin{array}{ccc} \mu_{1}&0&\\ 0&\mu_{2}&0\\0&0&\mu_{3} \end{array} \right) \mbox{ with real $\mu_{j}$'s}.\label{3.2b}
\end{eqnarray}
We can arbitrarily use the right-handed neutrino field $\nu'_{\rho R}$ given by $\nu'_{\rho R}(x)=\sum_{\sigma}W_{\rho\sigma}\nu_{\sigma R}(x), \mbox{ }W^{\dagger}W=I$.
While, the mass matrix $M'$(assumed to be $detM'\not = 0$) is uniquely expressed as
\begin{eqnarray}
M'=M\cdot U, \mbox{\hspace*{5mm}}UU^{\dagger}=I,\mbox{\hspace*{5mm}} M=[m_{\rho\sigma}],\label{3.3a}
\end{eqnarray}
where $M$ is hermitian as well as positive definite. (The last means all eigenvalue are positive.) 
Using the matrix $V$ which diagonalizes $M$ as
\begin{eqnarray}
V^{\dagger}MV=\left( \begin{array}{ccc} m_{1}&0&0\\ 0& m_{2}&0\\0&0&m_{3}\end{array} \right),
\mbox{\hspace*{5mm}} m_{j}>0, \mbox{\hspace*{5mm}}VV^{\dagger}=I,\label{3.3b}
\end{eqnarray}
we choose $W$ to be
\begin{eqnarray}
W=U^{\dagger};\label{3.3c}
\end{eqnarray}
then, by defining $\nu_{jL}$ and$ \nu_{jR}$ as
\begin{eqnarray}
\nu_{\sigma L/R}(x):=\sum_{j}V_{\sigma j}\nu_{j L/R}(x), \mbox{\hspace*{5mm}}V=[V_{\sigma j}],\label{3.3d}
\end{eqnarray}
the Lagrangian density(\ref{3.1}) is expressed as
\begin{eqnarray}
{\cal L}(x)&=&-\left( \begin{array}{ccc} \bar\nu_{eL}(x)&\bar\nu_{\mu L}(x)&\bar\nu_{\tau L}(x) \end{array} \right)(\not\partial + M)\left( \begin{array}{c} \nu_{eR}(x)\\ \nu_{\mu R}(x)\\ \nu_{\tau R}(x) \end{array} \right)-h.c.+{\cal L}_{int}(x) \nonumber \\
&=&-\left( \begin{array}{ccc} \bar\nu_{e}(x)&\bar\nu_{\mu}(x)&\bar\nu_{\tau}(x) \end{array} \right)(\not\partial + M)\left( \begin{array}{c} \nu_{e}(x)\\ \nu_{\mu}(x)\\ \nu_{\tau}(x) \end{array} \right)+{\cal L}_{int}(x);\label{3.3e}
\end{eqnarray}
the first term in the last line has the diagonal form, $-\sum^{3}_{j=1}\bar\nu_{j}(x)(\not\partial + m_{j})\nu_{j}(x).$

Similarly to (\ref{2.22}), we write the Hamiltonian density as
\begin{eqnarray}
{\cal H}(x)&=&{\cal H}^{0}(x)-{\cal L}_{int}(x),\label{3.5a}\nonumber\\ 
{\cal H}^{0}(x)&=&-\left( \begin{array}{ccc} \bar\nu_{e}(x)&\bar\nu_{\mu}(x)&\bar\nu_{\tau}(x) \end{array} \right)(\vec\gamma\vec\nabla +M^{0})\left( \begin{array}{c} \nu_{e}(x)\\ \nu_{\mu}(x)\\ \nu_{\tau}(x) \end{array} \right)+{\cal H}_{int}(x), \\
{\cal H}_{int}(x)&=&-\left( \begin{array}{ccc} \bar\nu_{e}(x)&\bar\nu_{\mu}(x)&\bar\nu_{\tau}(x) \end{array} \right)
\left( \begin{array}{ccc} \Delta_{ee}&m_{e\mu}&m_{e\tau}\\ m_{\mu e}&\Delta_{\mu\mu}&m_{\mu\tau}\\m_{\tau e}&m_{\tau\mu}&\Delta_{\tau\tau}\end{array} \right)
\left( \begin{array}{c} \nu_{e}(x)\\ \nu_{\mu}(x)\\ \nu_{\tau}(x) \end{array} \right),\label{3.5b}\\
M^{0}&=&\left( \begin{array}{ccc} m^{0}_{ee}&0&0\\ 0&m^{0}_{\mu\mu}&0\\0&0&m^{0}_{\tau\tau}\end{array} \right), \mbox{\hspace*{5mm}}\Delta_{\rho\rho}=m_{\rho\rho}-m^{0}_{\rho\rho}.\label{3.5c}
\end{eqnarray}

We give useful relations as follows.
\begin{eqnarray}
m_{\rho\sigma}&=&\sum^{3}_{j=1}v_{\rho j}\bar v_{\sigma j}m_{j}, \mbox{\hspace*{5mm}}\rho,\sigma=(e,\mu,\tau),\label{3.6a}\\
V^{\dagger}&=&V^{-1}=\frac{1}{detV}[u^{T}] \mbox{ with } u_{\sigma j}=\mbox{cofactor of } v_{\sigma j},\label{3.6b}\\
(detV)\sum_{j}v_{\rho j}\bar v_{\sigma j}&=&(detV)\delta_{\rho\sigma}=\sum_{j}v_{\rho j}u_{\sigma j},\label{3.6c}\\
(detV)\sum_{\rho}\bar v_{\rho j}v_{\rho k}&=&(detV)\delta_{jk}=\sum_{\rho}u_{\rho j}v_{\rho k},\label{3.6d}\\
\sum_{j}m_{j}&=&\sum_{\rho}m_{\rho\rho},\label{3.6e}\\
\Pi_{j}m_{j}&=&detM.\label{3.6f}
\end{eqnarray}
\subsection{Pole structure of Fourier transform of the neutrino propagator}We consider the Fourier transform of the neutrino propagator
\begin{eqnarray}
S'_{\sigma\rho}(\not k)=\mbox{Fourier trans. of}<0|T(\nu_{\sigma}(x)\bar\nu_{\rho}(y))|0>,
\label{3.7}
\end{eqnarray}
where $\nu_{\sigma}(x)$ and $\bar\nu_{\rho}(y)$ are the unrenormalized Heisenberg operator appearing in the Hamiltonian (\ref{3.5a}) with the interaction part ${\cal H}_{int}(x)-{\cal L}_{int}(x)$.

In the same way as described in the subsection B of Sec.\ref{sec2}, $S'_{\sigma\rho}(\not k)$ satisfies
\begin{eqnarray}
S'_{\sigma\rho}(\not k)=\delta_{\sigma\rho}S_{\rho}(\not k)+\sum_{\lambda}S'_{\sigma\lambda}(\not k)\Pi_{\lambda\rho}(\not k)S_{\rho}(\not k),\mbox{\hspace*{5mm}}\sigma, \rho=e,\mu,\tau,
\label{3.8}\end{eqnarray}
with $S_{\rho}(\not k)=(-\not k+im_{\rho\rho}^{0}+\epsilon)^{-1}$, and is expressed as
\begin{eqnarray}
S'_{\sigma\rho}(\not{k})=[f(\not{k})^{-1}]_{\sigma\rho}\mbox{ with }f_{\sigma\rho}(\not{k})=\delta_{\sigma\rho}S_{\rho}(\not{k})^{-1}-\Pi_{\sigma\rho}(\not{k}).
\label{3.9}\end{eqnarray}
Assuming the proper self-energy part $\Pi_{\sigma\rho}$ to be approximated as
\begin{eqnarray}
\Pi_{\sigma\rho} \simeq \Pi^{0}_{\sigma\rho}:=-i(M-M^{0})_{\sigma\rho},
\label{3.10a}\end{eqnarray}
we have
\begin{eqnarray}
[f_{\sigma\rho}(\not{k})]&=&[\delta_{\sigma\rho}(-\not k+im_{\rho\rho}^{0})+i(M-M^{0})_{\sigma\rho}]\nonumber\\
&=&\left( \begin{array}{ccc} -\not k+im_{ee}&im_{e\mu}&im_{e\tau}\\ im_{\mu e}&-\not k+im_{\mu\mu}&im_{\mu\tau}\\im_{\tau e}&im_{\tau\mu}&-\not k+im_{\tau\tau} \end{array} \right)
\label{3.10b}\end{eqnarray}
The physical one-particle masses are determined as three-poles obtained from
\begin{eqnarray}
det[f_{\sigma\rho}(\not{k})]=0.
\label{3.10c}\end{eqnarray}
From the form of (\ref{3.10b}), we see that the arbitrariness in separating ${\cal H}_{int}(x)$ from the "free" part in (\ref{3.5b}), i.e. the arbitrariness in defining $S_{\rho}(\not k)$, disappears in the physical one-particle masses under the approximation (\ref{3.10a}).
These one-particle masses determined from (\ref{3.10c}) with $f_{\sigma\rho}(\not k)$ given by (\ref{3.10b}) coincide with the eigenvalues $\{m_{j}, j=1,2,3\}$ of the mass matrix $M=[m_{\rho\sigma}]$.
\subsection{Diagonalization of pole part in the propagator}\label{sec3.3}
We follow the same procedure of the diagonalization as described in the subsection C of Sec.\ref{sec2}.
Writing the cofactor of $f_{\sigma\rho}(\not k)$ as $F_{\sigma\rho}(\not k)$, we write $S'_{\sigma\rho}(\not k)$ in the same form as (\ref{2.26}); then, we obtain
\begin{eqnarray}
det[f_{\sigma\rho}]|_{\not k=im_{j}}=0=[\sum_{\lambda}f_{\sigma\lambda}^{(j)}F_{\rho\lambda}^{(j)}]=[F_{\lambda\sigma}^{(j)}f_{\lambda\rho}^{(j)}].
\label{3.11}\end{eqnarray}
The explicit form of $\rho^{(j)}$ defined in the same way as (\ref{2.28}) is written as
\begin{eqnarray}
(\rho^{(j)})^{-1}=-(m_{ee}-m_{j})(m_{\mu\mu}-m_{j})-(m_{\mu\mu}-m_{j})(m_{\tau\tau}-m_{j})-(m_{\tau\tau}-m_{j})(m_{ee}-m_{j})\nonumber\\
\mbox{\hspace*{\fill}}+m_{e\mu}m_{\mu e}+m_{\mu\tau}m_{\tau\mu}+m_{e\tau}m_{\tau e}.
\label{3.12a}\end{eqnarray}
By employing (\ref{3.6a}) and (\ref{3.6b}), we obtain after some calculations
\begin{eqnarray}
(\rho^{(1)})^{-1}&=&-(m_{1}-m_{2})(m_{1}-m_{3}),\nonumber\\
(\rho^{(2)})^{-1}&=&-(m_{2}-m_{1})(m_{2}-m_{3}),\label{3.12b}\\
(\rho^{(3)})^{-1}&=&-(m_{3}-m_{1})(m_{3}-m_{2}).\nonumber
\end{eqnarray}

As to $F_{\rho\tau}^{(j)}$'s, expressed from the definition as
\begin{eqnarray}
F^{(j)}_{e\tau}&=&-m_{\mu e}m_{\tau\mu}+m_{\tau e}(m_{\mu\mu}-m_{j}),\nonumber\\
F^{(j)}_{\mu\tau}&=&-m_{e\mu}m_{\tau e}+m_{\tau\mu}(m_{ee}-m_{j}),\label{3.13a}\\
F^{(j)}_{\tau\tau}&=&m_{\mu e}m_{e\mu}-(m_{ee}-m_{j})(m_{\mu\mu}-m_{j}),\nonumber
\end{eqnarray}
some calculations lead to
\begin{eqnarray}
F_{\rho\tau}^{(j)}=\frac{\bar v_{\rho j}v_{\tau j}}{\rho^{(j)}}, \mbox{\hspace*{5mm}}\rho=e, \mu,\tau,\mbox{\hspace*{5mm}} j=1,2,3;
\label{3.13b}\end{eqnarray}
thus, we obtain
\begin{eqnarray}
\frac{\rho^{(j)}}{F^{(j)}_{\tau\tau}}=\frac{(\rho^{(j)}) ^{2}}{|v_{\tau j}|^{2}}.
\label{3.13c}\end{eqnarray}

Next we define a set of new fields $\psi^{r}_{j}(x)$ and $\bar\psi^{r}_{j}(x), j=1,2,3$, in the same way as (\ref{2.31}).
The condition for determining the matrix $A$ is
\begin{eqnarray}
\rho^{(j)}(F^{(j)})^{T}=AE^{(j)}A^{\dagger}
\label{3.14}\end{eqnarray}
with
\begin{eqnarray}
E^{(1)}=\left( \begin{array}{ccc} 1&0&0\\ 0&0&0\\ 0&0&0 \end{array} \right),\mbox{\hspace*{5mm}}E^{(2)}=\left( \begin{array}{ccc} 0&0&0\\ 0&1&0\\ 0&0&0 \end{array} \right),\mbox{\hspace*{5mm}}E^{(3)}=\left( \begin{array}{ccc} 0&0&0\\ 0&0&0\\ 0&0&1 \end{array} \right).\nonumber
\end{eqnarray}
The above equation leads to
\begin{eqnarray}
\bar A_{\rho j}A_{\sigma j}=\rho^{(j)}F_{\rho\sigma}^{(j)}=\frac{\bar A_{\rho j}A_{\tau j}\bar A_{\tau j}A_{\sigma j}}{\bar A_{\tau j}A_{\tau j}}=\frac{\rho^{(j)}\bar F_{\tau\rho}^{(j)}F_{\tau\sigma}^{(j)}}{F_{\tau\tau}^{(j)}};
\label{3.15a}\end{eqnarray}
therefore, noting (\ref{3.13c}) we are allowed to take
\begin{eqnarray}
A_{\sigma j}=\sqrt{\frac{\rho^{(j)}}{F_{\tau\tau}^{(j)}}}F_{\tau\sigma}^{(j)}\omega=\frac{\omega|\rho^{(j)}|}{|v_{\tau j}|}F_{\tau\sigma}^{(j)}, \mbox{\hspace*{5mm}}|\omega|^{2}=1.
\label{3.15b}\end{eqnarray}
Employing the concrete forms (\ref{3.13b}) of $F_{\tau\sigma}^{(j)}=\bar F_{\sigma\tau}^{(j)}$, we obtain 
\begin{eqnarray}
\left( \begin{array}{c} A_{e1}\\ A_{\mu 1}\\ A_{\tau 1} \end{array} \right)=\frac{\omega|\rho^{(1)}|\bar v_{\tau 1}}{\rho^{(1)}|v_{\tau 1}|}\left( \begin{array}{c} v_{e1}\\ v_{\mu 1}\\ v_{\tau 1} \end{array} \right)=\epsilon_{1}\left( \begin{array}{c} v_{e1}\\ v_{\mu 1}\\ v_{\tau 1} \end{array} \right),\\
\left( \begin{array}{c} A_{e2}\\ A_{\mu 2}\\ A_{\tau 2} \end{array} \right)=\frac{\omega|\rho^{(2)}|\bar v_{\tau 2}}{\rho^{(2)}|v_{\tau 2}|}\left( \begin{array}{c} v_{e2}\\ v_{\mu 2}\\ v_{\tau 2} \end{array} \right)=\epsilon_{2}\left( \begin{array}{c} v_{e2}\\ v_{\mu 2}\\ v_{\tau 2} \end{array} \right),\\
\left( \begin{array}{c} A_{e3}\\ A_{\mu 3}\\ A_{\tau 3} \end{array} \right)=\frac{\omega|\rho^{(3)}|\bar v_{\tau 3}}{\rho^{(3)}|v_{\tau 3}|}\left( \begin{array}{c} v_{e3}\\ v_{\mu 3}\\ v_{\tau 3} \end{array} \right)=\epsilon_{3}\left( \begin{array}{c} v_{e3}\\ v_{\mu 3}\\ v_{\tau 3} \end{array} \right),
\label{3.16}\end{eqnarray}
where 
\begin{eqnarray}
\epsilon_{j}:=\frac{\omega|\rho^{(j)}|\bar v_{\tau j}}{\rho^{(j)}|v_{\tau j}|}.
\label{3.16d}\end{eqnarray}
By choosing the order as $m_{3}>m_{2}>m_{1}$, we have
\begin{eqnarray}
\rho^{(1)}<0, \mbox{\hspace*{5mm}}\rho^{(2)}>0, \mbox{\hspace*{5mm}}\rho^{(3)}<0;
\label{3.17a}\end{eqnarray}
then,
\begin{eqnarray}
\epsilon_{j}=(-1)^{j}\frac{\omega v_{\tau j}}{|\bar v_{\tau j}|}.
\label{3.17b}\end{eqnarray}
Thus the form of the matrix $A$ satisfies the unitary condition, i.e. 
\begin{eqnarray}
AA^{\dagger}=A^{\dagger}A=I,
\label{3.18a}\end{eqnarray}
and is essentially the same as $V$ which diagonalizes the mass matrix $M$;
\begin{eqnarray}
A=[A_{\rho j}]&=&V\cdot E \mbox{\mbox{\hspace*{5mm}} with \mbox{\hspace*{5mm}}} E=\left( \begin{array}{ccc} \epsilon_{1}&0&0\\ 0&\epsilon_{2}&0\\ 0&0&\epsilon_{3} \end{array} \right),\\
A^{\dagger}MA&=&\left( \begin{array}{ccc} m_{1}&0&0\\ 0&m_{2}&0\\ 0&0&m_{3} \end{array} \right).
\label{3.18}\end{eqnarray}
$S^{r}_{ij}(\not k)$, the Fourier transform of $<0|T(\psi^{r}_{i}(x)\bar\psi^{r}_{j}(y))|0>$, is now written as
\begin{eqnarray}
[S^{r}_{ij}(\not k)]&=&A^{\dagger}S'(\not k)A\nonumber\\
&=&[\sum_{l}\frac{1}{-\not k+im_{1}+\epsilon}(\sum_{\sigma\rho}\bar A_{\sigma i}\rho^{(l)}F_{\sigma\rho}^{(l)}A_{\rho j})]+\mbox{(contribution from continuous spectra)}\nonumber\\
&=&\left( \begin{array}{ccc} \frac{1}{-\not k+im_{1}+\epsilon}&0&0\\ 0& \frac{1}{-\not k+im_{2}+\epsilon}&0\\ 0&0& \frac{1}{-\not k+im_{3}+\epsilon} \end{array} \right)+\mbox{(contribution from continuous spectra)},
\label{3.19}\end{eqnarray}
and as to $S'_{\sigma\rho}(\not k)$ we obtain the same fom as given by (\ref{2.44}).
\section{Comments on the choice (1.4)}\label{sec4}

We examine the problem whether or not there is any compelling reason for choosing (\ref{1.4}), which is expressed in a convenient form for the following consideration as
\begin{eqnarray}
\left( \begin{array}{c} \tilde\alpha_{\sigma}(kr;t)\\\tilde\beta^{\dagger}_{\sigma}(-kr;t)\end{array} \right)=G^{-1}(\theta;t)\left( \begin{array}{c} \alpha_{j}(kr;t)\\ \beta^{\dagger}_{j}(-kr;t) \end{array} \right)G(\theta;t),\mbox{\hspace*{5mm}} (\sigma, j)=(e,1), (\mu, 2);
\label{4.1}\end{eqnarray}
here $\tilde\beta_{j}(-k,r;t)=\tilde\beta_{j}(q,r;t)$ with $\vec q=-\vec k$ and $q_{0}=\sqrt{\vec k^{2}+m_{j}^{2}}=\omega_{j}$(here we use the notations $(\tilde\alpha_{\sigma}, \tilde\beta_{\sigma})$ instead of $(\alpha^{BV}_{\sigma}, \beta^{BV}_{\sigma})$ in (\ref{1.4})).
As noted in Sec.\ref{sec2}, when neglecting ${\cal L}_{int}(x)$ in the Lagrangian (\ref{2.1}), the diagonalization of the one-pole term in the propagator corresponds to the diagonalization of the mass-term in (\ref{2.1}) through
\begin{eqnarray}
\left( \begin{array}{c} \nu_{e}(x)\\ \nu_{\mu}(x) \end{array} \right)=G^{-1}(\theta;t)\left( \begin{array}{c} \nu_{1}(x)\\ \nu_{2}(x) \end{array} \right)G(\theta;t))=\left( \begin{array}{cc} c_{\theta}&s_{\theta}\\ -s_{\theta}&c_{\theta} \end{array} \right)\left( \begin{array}{c} \nu_{1}(x)\\ \nu_{2}(x) \end{array} \right),
\label{4.2}\end{eqnarray}
and $\nu_{j}(x)$ is expanded as
\begin{eqnarray}
\nu_{j}(x)=\frac{1}{\sqrt{V}}\sum_{\vec k r}e^{i\vec k\vec x}\{u_{j}(k,r)\alpha_{j}(k,r;t)+v_{j}(-k,r)\beta^{\dagger}_{j}(-k,r;t)\}
\label{4.3}\end{eqnarray}
with
\begin{eqnarray}
\left( \begin{array}{c} \alpha_{j}(kr;t)\\ \beta_{j}(kr;t) \end{array} \right)=\left( \begin{array}{c} \alpha_{j}(kr;0)\\ \beta_{j}(kr;0) \end{array} \right)e^{-i\omega_{j}t}, \mbox{\hspace*{5mm}}\omega_{j}=(\vec k^{2}+m^{2}_{j})^{\frac{1}{2}}
\label{4.4}\end{eqnarray}
If (\ref{4.1}) is allowed to think to define the creation and  annihilation operators for definte flavor states, (\ref{4.2}) and (\ref{4.3}) lead to the expansion
\begin{eqnarray}\nu_{\sigma}(x)=\frac{1}{\sqrt{V}}\sum_{\vec k r}e^{i\vec k\vec x}\{u_{j}(k,r)\tilde\alpha_{\sigma}(k,r;t)+v_{j}(-k,r)\tilde\beta^{\dagger}_{\sigma}(-k,r;t)\}.\label{4.6}\end{eqnarray}
The definition of (\ref{4.1}), employed in refs.\cite{BV} and \cite{BHV}, is certainly the simplest one which is consistent with (\ref{4.2}); then, (\ref{4.1}) is expressed in terms of the $\nu$-fields as
\begin{eqnarray}\left( \begin{array}{c} \tilde\alpha_{\sigma}(kr;t)\\\tilde\beta^{\dagger}_{\sigma}(-kr;t)\end{array} \right)=\frac{1}{\sqrt{V}}\int d^{3}xe^{-i\vec k\vec x}\left( \begin{array}{c}
u_{j}^{\dagger}(kr)\\ v_{j}^{\dagger}(-kr) \end{array} \right)G^{-1}(\theta;t)\nu_{j}(x)G(\theta;t).
\label{4.7}\end{eqnarray}
We cannot exclude, however, other ways; really, we can define generally
\begin{eqnarray}
\left( \begin{array}{c} \alpha_{\sigma}(kr;t)\\ \beta^{\dagger}_{\sigma}(-kr;t)\end{array} \right)&=&\frac{1}{\sqrt{V}}\int d^{3}xe^{-i\vec k\vec x}\left( \begin{array}{c} u_{\sigma}^{\dagger}(kr)\\ v_{\sigma}^{\dagger}(-kr)
\end{array} \right)G^{-1}(\theta;t)\nu_{j}(x)G(\theta;t)\nonumber \\
&=&G^{-1}(\theta;t)\left( \begin{array}{c} \rho_{\sigma j}(k)\alpha_{j}(k,r;t)+i\lambda_{\sigma j}(k)\beta^{\dagger}_{j}(-k,r;t)\\ i\lambda_{\sigma j}(k)\alpha_{j}(k,r;t)+ \rho_{\sigma j}(k)\beta^{\dagger}_{j}(-k,r;t)\end{array} \right)G(\theta;t)
\label{4.8}\end{eqnarray}
for an arbitrary $m_{\sigma}$; 
(\ref{4.2}) leads to the expansion
\begin{eqnarray}\nu_{\sigma}(x)=\frac{1}{\sqrt{V}}\sum_{\vec k r}e^{i\vec k\vec x}\{u_{\sigma}(k,r)\alpha_{\sigma}(k,r;t)+v_{\sigma}(-k,r)\beta^{\dagger}_{\sigma}(-k,r;t)\}.
\label{4.9}\end{eqnarray}
(See Appendix A as to the definitions of $\rho_{\sigma j}$ and $\lambda_{\sigma j}$.) 
(\ref{4.8}) leads to the transformation (\ref{1.9}).

For the purpose of finding any logical basis of the choice (\ref{4.1}), we reexamine the consideration in ref.\cite{BV}. We introduce the mass and the flavor vacua $|0>_{m}$ and $|0(\theta,t)>$ as
\begin{eqnarray}
\left\{ \begin{array}{c} \alpha_{j}(kr;t)\\ \beta_{j}(-kr;t)\end{array} \right\}|0>_{m}=0,\mbox{\hspace*{5mm}}\left\{ \begin{array}{c} \alpha_{\sigma}(kr;t)\\ \beta_{\sigma}(-kr;t)\end{array} \right\}|0(\theta ;t)>=0
\label{4.10}\end{eqnarray}
for $^{\forall}\vec k, r, j$ and $\sigma$. 
It should be remembered that the authors of ref.\cite{BV} choose the special vacuum as $|0(\theta,t)>$, given(at finite volume $V$) by
\begin{eqnarray}
|0(\theta ;t)>=G^{-1}(\theta ;t)|0>_{m},\mbox{\hspace*{5mm}} _{m}<0|0>_{m}=1.
\label{4.11}\end{eqnarray}
The reasoning why the relation (\ref{4.11}) is adoted in ref.\cite{BV} is as follows.
For $^{\forall}|a>_{m}$ and $|b>_{m}\in {\cal H}_{m}$, which is the Fock space constructed in terms of the $\nu_{j}$-fields, we have
\begin{eqnarray}
_{m}<a|G(\theta ;t)\nu_{\sigma}(x)G^{-1}(\theta ;t)|b>_{m}=_{m}<a|\nu_{j}(x)|b>_{m};
\label{4.12}\end{eqnarray}
$G^{-1}(\theta ;t)|b>_{m}$ should belong to the flavor Fock space ${\cal H}_{f}$ constructed in terms of the $\nu_{\sigma}$-fields and gives the mapping of ${\cal H}_{m}$ to ${\cal H}_{f}$; especially we obtain (\ref{4.11}).

Then, by operating $\alpha_{\sigma}(k,r;t)$ to $|0(\theta ;t)>$, we obtain from (\ref{4.8}) and (\ref{4.11}) the constraint
\begin{eqnarray}
\lambda_{\sigma j}(k)=0, \mbox{ i.e. } \rho_{\sigma j}(k)=1 \mbox{ for } ^{\forall}k \mbox{ and } (\sigma, j)=(e, 1),(\mu, 2),
\label{4.13}\end{eqnarray}
which leads us to take $m_{e}=m_{1}$ and $m_{\mu}=m_{2}$.
Also, by operating $\beta^{\dagger}_{\sigma}(-k,r;t)$ to $|0(\theta ;t)>$, the same constraint as (\ref{4.13}) is obtained from the norms of 
\begin{eqnarray}
\beta^{\dagger}_{\sigma}(-k,r;t)|0(\theta ;t)>=G^{-1}(\theta ;t)\rho_{\sigma j}(k)\beta^{\dagger}_{j}(-k,r;t)|0>_{m}.
\label{4.14}\end{eqnarray}
In this way, the simplest choice (\ref{4.1}) is derived.

It seems necessary, however, for us to reconsider the content of deriving (\ref{4.13}).
The general relation (\ref{4.8}) is rewritten as
\begin{eqnarray}
\left( \begin{array}{c} \alpha_{\sigma}(kr;t)\\ \beta^{\dagger}_{\sigma}(-kr;t)\end{array} \right)&=&\left( \begin{array}{cc} \rho_{\sigma j}(k)&i\lambda_{\sigma j}(k)\\ i\lambda_{\sigma j}(k)&\rho_{\sigma j}(k) \end{array} \right)\left( \begin{array}{c} \tilde\alpha_{\sigma}(kr;t)\\ \tilde\beta^{\dagger}_{\sigma}(-kr;t)\end{array} \right)\nonumber\\
&=&I^{-1}(t)\left( \begin{array}{c} \tilde\alpha_{\sigma}(kr;t)\\ \tilde\beta^{\dagger}_{\sigma}(-kr;t)\end{array} \right)I(t)
\label{4.15}\end{eqnarray}
where
\begin{eqnarray}
I(t)=\prod_{\vec k, r}exp\{i\sum_{(\sigma, j)}\xi_{\sigma, j}(k)(\tilde\alpha^{\dagger}_{\sigma}(k,r;t)\tilde\beta^{\dagger}_{\sigma}(-k,r;t)+\tilde\beta_{\sigma}(-k,r;t)\tilde\alpha_{\sigma}(k,r;t))\}
\label{4.16}\end{eqnarray}
with $cos\xi_{\sigma, j}(k)=\rho_{\sigma j}(k)=cos\frac{\chi_{\sigma}-\chi_{j}}{2}$.
In (\ref{4.16}), the summation $\sum_{(\sigma, j)}$ means to take the sum over the two sets, $(e, 1)$ and $(\mu, 2)$. 
Thus one can construct Hilbert spaces ${\cal H}_{m}$ and ${\cal H}_{f}$ by operationg possible polynomials of $\{\alpha^{\dagger}_{j}\mbox{'s}, \mbox{\hspace*{3mm}}\beta^{\dagger}_{j}\mbox{'s}, \mbox{\hspace*{3mm}}j=1,2\}$ and $\{\alpha^{\dagger}_{\sigma}\mbox{'s}, \mbox{\hspace*{3mm}} \beta^{\dagger}_{\sigma}\mbox{'s}, \mbox{\hspace*{3mm}} \sigma=e, \mu\}$, respectively, on the vacuum states $|0>_{m}$ and $|0(\theta ;t)>$, where these state are defined by
\begin{eqnarray}
\alpha_{j}(k,r;t)|0>_{m}&=&\beta_{j}(k,r;t)|0>_{m}=0 \mbox{\hspace*{3mm} for\hspace*{3mm} } ^{\forall}j, \vec k, r,
\label{4.17}\\
\alpha_{\sigma}(k,r;t)|0(\theta;t)>&=&\beta_{\sigma}(k,r;t)|0(\theta;t)>=0 \mbox{\hspace*{3mm} for\hspace*{3mm} } ^{\forall}\sigma, \vec k, r.
\label{4.18}\end{eqnarray}
Since (\ref{4.15}) is expressed as 
\begin{eqnarray}
\left( \begin{array}{c} \alpha_{\sigma}(kr;t)\\ \beta^{\dagger}_{\sigma}(kr;t) 
\end{array} \right)=(G(\theta;t)I(t))^{-1}\left( \begin{array}{c} \alpha_{j}(kr;t)\\ \beta^{\dagger}_{j}(kr;t) \end{array} \right)(G(\theta;t)I(t)).
\label{4.19}\end{eqnarray}
We see, after repeating the same argument as in ref.\cite{BV} explained above, that the relation between the vacuum states is given (at finite volume) by
\begin{eqnarray}
|0(\theta ;t)>=(G(\theta ;t)I(t))^{-1}|0>_{m};
\label{4.20}\end{eqnarray}
thus one cannot obtain any constraint on $m_{\sigma}$'s.
Therefore, the choice (\ref{4.1}) has no compelling theoretical reason other than "convenience" and "simplicity" and there is no physical basis for fixing the $m_{\sigma}$ values.

\section{Summary and final remarks}\label{sec5}

We have first examined the diagonalizaion of the flavor-neutrino propagator matrix by following the procedure proposed by Kaneko et.al.\cite{kaneko} (but without employing any intermediate fields $\{\phi_{j}, \tilde\phi_{j}, j=1,2,3\}$).
We have concretely shown that, in so far as the matrix of the proper self-energy part $\Pi_{\rho\sigma}({\not k})$ for the flavor-neutrino fields is allowed to be approximated by neglecting ${\cal L}_{int}$ in (\ref{3.3e}), a set of the renormalized fields $\{\psi_{j}^{r}(x),\bar\psi_{j}^{r}(x)\}$ can be defined as
\begin{eqnarray}
\psi_{\sigma}(x)=\sum_{j=1}^{3}z^{\frac{1}{2}}_{\sigma j}\psi_{j}^{r}(x), \mbox{\hspace*{5mm}}\bar\psi_{\sigma}(x)=\sum_{j=1}^{3}\bar z^{\frac{1}{2}}_{\sigma j}\bar \psi_{j}^{r}(x), \mbox{\hspace*{5mm}}\sigma=e,\mu,\tau
\label{5.1}\end{eqnarray}
with ${z^{\frac{1}{2}}}^{\dagger}z^{\frac{1}{2}}=I$, so that the Fourier transform of $<0|T(\psi^{r}_{i}(x)\bar\psi^{r}_{j}(y))|0>$, i.e. $S_{ij}^{r}(\not k)$, has a single one-pole term, and
\begin{eqnarray}
S'_{\sigma\rho}(\not k)&=&\mbox{ F.T. of }<0|T(\psi_{\sigma}(x)\bar\psi_{\rho}(y))|0>\nonumber\\
&=&\sum_{j=1}^{3}\frac{z^{\frac{1}{2}}_{\sigma j}\bar z^{\frac{1}{2}}_{\rho j}}{-\not k +im_{j}+\epsilon}+\int d(\kappa^{2})\frac{\lambda_{\sigma\rho}(\kappa,\not k)}{k^{2}+\kappa^{2}-i\epsilon}.
\label{5.2}\end{eqnarray}
The matrix $[z^{\frac{1}{2}}]$, which may be called the generalized z-factor, has been shown to be essentially the same as that diagonalizing the mass matrix $M$ in the starting Lagrangian (\ref{3.3e}).

Under the adopted approximation for $\Pi_{\sigma\rho}$, the content summarized above seems consistent and, in some sense self-evident.
We cannot go, however, beyond this approximation due to ignorance of the Higgs neutrino interaction included in ${\cal L}_{int}(x)$.

Under the same approximation we have reexamined in Sec.\ref{sec4} the problem proposed by Blasone et al. \cite{BV}.
The essential problem in the field theory of the neutrino mixing is to settle how to define appropriate creation and annihilation operators of the flavor(or weak\cite{giunti}) states.
We have shown that, by taking account of the general relation (\ref{4.19}) as well as the relation between the two vacuum states, (\ref{4.20}), there is neither theoretical reason for choosing $m_{e}=m_{1}$ and $m_{\mu}=m_{2}$ adopted in ref.\cite{BV} nor physical basis for fixing any special $m_{\sigma}$ values.
In this sense we cannot construct generally the flavor Fock space, except for the extremely relativistic case; this is in accordance with the assertion in ref.\cite{giunti}, though our reasoning is based on a different context.

In so far as we consider the transition or survival amplitudes by treating such a quantity as
\begin{eqnarray}
{\cal A}_{\rho\sigma}(k,r;t)=_{m}<0|\alpha_{\rho}(k,r;t)\alpha^{\dagger}_{\rho}(k,r;0)|0>_{m}
\label{5.3}\end{eqnarray}
we have to fix the mass parameters $m_{\rho}$'s in order to settle the relations of $\alpha_{\rho}$'s to the fields $\nu_{\rho}(x)$'s.
But we saw in Sec.\ref{sec4} that there is no theoretical reason to specify the mass parameters.
Instead, under the approximation of neglecting ${\cal L}_{int}$ in the Lagrangian (\ref{3.3e}), the propagator, $S'_{\sigma\rho}(k)=\mbox{F.T. of }<0|T(\nu_{\sigma}(x)\bar\nu_{\rho}(y))|0>$, has been shown to have a structure which is independent of the mass parameter $m_{\rho}$'s, that is, independent of the choice of the perturbative vacuum corresponding to the 'free' Hamiltonians, specified by the mass parameters $m^{0}_{\rho\rho}$'s as in (\ref{2.22}) and (\ref{3.5a}).
Thus, it is favorable for us to treat the neutrino oscillation problem by relying solely on the neutrino propagator.
Related investigations have been done in ref.\cite{rich}, and the work developed by Grimus and Stockinger\cite{grimus} seems to be important from our viewpoint.
\acknowledgments
The authors would like to expressed their thanks to Prof. K. Ishikawa for various suggestions.
Thanks are also due to Prof. S. M. Bilenky for his remark during the satellite symposium on neutrino, "New Era in Neutrino Physics" held at Tokyo Metropolitan University, June 1998.
One of the authors, T.Y., would like to thank to Prof. G. Vitiello, Dr. M. Blasone and Dr. E. Alfinito for dicussions as well as for the hospitality during T.Y.'s stay in Salerno University.


\appendix
\section{}
Explicit forms of the plane-wave eigenfunctions $u(kr)$ and $v(kr)$, satisfying
\begin{eqnarray}
(i\not k +m)u(kr)=0, \mbox{\hspace*{5mm}}(-i\not k +m)v(kr)=0
\label{a.1}\end{eqnarray}
are given, in the Kramers representation of $\gamma$-matrices (i.e. $\vec\gamma =-\rho_{y}\otimes \vec\sigma, \gamma^{4}=\rho_{x}\otimes I, \gamma_{5}=-\rho_{z}\otimes I$)\cite{cornaldesi}, by
\begin{eqnarray}
u(k\uparrow)=\left( \begin{array}{c} c\alpha\\ c\beta\\ s\alpha\\ s\beta \end{array} \right), \mbox{\hspace*{5mm}} u(k\downarrow)=\left( \begin{array}{c} -s\beta^{*}\\s\alpha^{*}\\ -c\beta^{*} \\c\alpha^{*}  \end{array} \right), \mbox{\hspace*{5mm}}\label{a.2}\\
v(k\uparrow)=\left( \begin{array}{c} s\beta^{*}\\-s\alpha^{*}\\ -c\beta^{*} \\c\alpha^{*}  \end{array} \right), \mbox{\hspace*{5mm}}v(k\downarrow)=\left( \begin{array}{c} c\alpha\\ c\beta\\ -s\alpha\\ -s\beta \end{array} \right).
\label{a.3}\end{eqnarray}
Here, $c=cos(\frac{\chi}{2}), s=sin(\frac{\chi}{2}), cot\chi =\frac{|\vec k|}{m}, k_{z}=kcos\vartheta, k_{x}+ik_{y}=ksin\vartheta\cdot e^{i\phi}, \alpha=cos(\frac{\vartheta}{2})\cdot e^{-i\phi /2}, \beta=sin(\frac{\vartheta}{2})\cdot e^{i\phi /2}$. 
$u(kr)$ and $v(kr)$ are the eigenfunctions of the helicity $\vec s\cdot \vec k /|\vec k|, \:\vec s=(I\times \vec\sigma)/2$;
\begin{eqnarray}
\frac{1}{k}(\vec s\cdot \vec k)u(k\uparrow)=\frac{1}{2}u(k\uparrow), \mbox{\hspace*{5mm}}\frac{1}{k}(\vec s\cdot \vec k)u(k\downarrow)=-\frac{1}{2}u(k\downarrow), \nonumber\\
\frac{1}{k}(\vec s\cdot -\vec k)v(k\uparrow)=\frac{1}{2}v(k\uparrow), \mbox{\hspace*{5mm}}\frac{1}{k}(\vec s\cdot -\vec k)v(k\downarrow)=-\frac{1}{2}u(k\downarrow).
\label{a.4}\end{eqnarray}
The solutions of (\ref{a.1}) with the mass $m_{j}$ are written as $u_{j}(kr)$ and $v_{j}(kr)$.
We obtain
\begin{eqnarray}
u_{1}^{*}(kr)u_{2}(ks)=v_{1}^{*}(-kr)v_{2}(-ks)=\rho_{12}(k),\nonumber\\
u_{1}^{*}(kr)v_{2}(-ks)=v_{1}^{*}(-kr)u_{2}(ks)=i\lambda_{12}(k),
\label{a.5}\end{eqnarray}
where $v_{j}(-kr):=v(pr)$ with $\vec p=-\vec k, \:p_{0}=k_{0j}=\sqrt{\vec k^{2}+m_{j}^{2}}$;
$\rho_{12}=cos\frac{\chi_{1}-\chi_{2}}{2}, \:\lambda_{12}=sin\frac{\chi_{1}-\chi_{2}}{2}$ with $cot\chi_{j}=\frac{|\vec k|}{m_{j}}$.
We have
\begin{eqnarray}
\sum_{r}\{u^{b}_{j}(kr)\cdot u^{d}_{j}(kr)^{*}+v^{b}_{j}(-kr)\cdot v^{d}_{j}(-kr)^{*}\}=\delta_{bd}.
\label{a.6}\end{eqnarray}
The explicit form of ${\cal G}(\theta,k)$ appearing in (\ref{1.9}) is given by
\begin{eqnarray}
{\cal G}(\theta,k)=\left( \begin{array}{cc} P(\theta,k)&i\Lambda(\theta,k)\\ i\Lambda(\theta,k)&P(\theta,k) \end{array} \right)
\label{a.7}\end{eqnarray}
with
\begin{eqnarray}
P(\theta,k)=\left( \begin{array}{cc} c_{\theta}\rho_{e1}(k)&s_{\theta}\rho_{e2}(k)\\ -s_{\theta}\rho_{\mu 1}(k)&c_{\theta}\rho_{\mu 2}(k) \end{array} \right),\mbox{\hspace*{5mm}} \Lambda(\theta,k)=\left( \begin{array}{cc} c_{\theta}\lambda_{e1}(k)&s_{\theta}\lambda_{e2}(k)\\ -s_{\theta}\lambda_{\mu 1}(k)&c_{\theta}\lambda_{\mu 2}(k) \end{array} \right).\label{a.8}
\end{eqnarray}
${\cal G}(\theta,k)$ is confirmed to be unitary;
\begin{eqnarray}
{\cal G}(\theta,k){\cal G}(\theta,k)^{\dagger}={\cal G}(\theta,k)^{\dagger}{\cal G}(\theta,k)=I.
\label{a.9}\end{eqnarray}
\section{}
Here we will give a remark on ref.\cite{giunti} as follows.
We define
\begin{eqnarray}
w(k\uparrow):=\left[\begin{array}{c} \alpha\\ \beta \end{array}\right ], \mbox{\hspace*{5mm}}w(k\downarrow):=\left[\begin{array}{c} -\beta^{*}\\ \alpha^{*} \end{array}\right] \mbox{ with } \alpha \mbox{ and } \beta \mbox{ used in App.A}, 
\label{b.1}\end{eqnarray}
which satisfy
\begin{eqnarray}
w(k\uparrow)^{\dagger}w(k\downarrow)=0, \mbox{\hspace*{5mm}}\frac{\vec\sigma\vec k}{k}w(kr)=w(kr)\left\{ \begin{array}{c} 1\\ -1 \end{array} \right\}\mbox{ for }\left\{\begin{array}{c} r=\uparrow\\r=\downarrow  \end{array}\right\}.
\label{b.2}\end{eqnarray}
$u_{j}(kr)$ and $v_{j}(-kr)$ given by (\ref{a.2}) and (\ref{a.3}) are expressed as
\begin{eqnarray}
u(k\uparrow)&=&\left[ \begin{array}{c} \kappa_{+}\\ \kappa_{-} \end{array} \right]\otimes w(k\uparrow), \mbox{\hspace*{5mm}}u(k\downarrow)=\left[ \begin{array}{c} \kappa_{-}\\ \kappa_{+} \end{array} \right]\otimes w(k\downarrow),\nonumber\\
v(-k\uparrow)&=&i\left[ \begin{array}{c} -\kappa_{-}\\ \kappa_{+} \end{array} \right]\otimes w(k\uparrow), \mbox{\hspace*{5mm}}v(-k\downarrow)=i\left[ \begin{array}{c} \kappa_{+}\\ -\kappa_{-} \end{array} \right]\otimes w(k\downarrow),
\label{b.3}\end{eqnarray}
where $\kappa_{\pm}:=\sqrt{(E\pm k)/(2E)}=\{\begin{array}{c} c\\ s \end{array}\}, E=\sqrt{k^{2}+m^{2}}$.
By using the unitary matrix $U=(U_{\sigma j})$ connecting the two kinds of neutrino fields, we obtain
\begin{eqnarray}
\nu_{\sigma}(x)=\frac{1}{\sqrt{V}}\sum_{\vec k}\sum_{j}\left(
U_{\sigma j}\alpha_{j}(k\uparrow;t)\left[ \begin{array}{c} \kappa_{j+}\\ \kappa_{j-} \end{array} \right]\otimes w(k\uparrow)+
U_{\sigma j}\alpha_{j}(k\downarrow;t)\left[ \begin{array}{c} \kappa_{j-}\\ \kappa_{j+} \end{array} \right]\otimes w(k\downarrow)\right. \nonumber\\
\left. +iU_{\sigma j}\beta_{j}^{\dagger}(-k\uparrow;t)\left[ \begin{array}{c} -\kappa_{j-}\\ \kappa_{j+} \end{array} \right]\otimes w(k\uparrow)
+iU_{\sigma j}\beta_{j}^{\dagger}(-k\downarrow;t)\left[ \begin{array}{c} \kappa_{j+}\\ -\kappa_{j-} \end{array} \right]\otimes w(k\downarrow)\right)e^{i\vec k\vec x}.
\label{b.4}\end{eqnarray}
By defining
\begin{eqnarray}
A_{\sigma \pm}(kr;t)&:=&\sum_{j}U_{\sigma j}\alpha_{j}(kr;t)\kappa_{j\pm},\nonumber\\
B^{\dagger}_{\sigma \pm}(-kr;t)&:=&\sum_{j}U_{\sigma j}\beta^{\dagger}_{j}(-kr;t)\kappa_{j\pm}(\pm i),
\label{b.5}\end{eqnarray}
(\ref{b.4}) is rewritten as
\begin{eqnarray}
\nu_{\sigma}(x)=\frac{1}{\sqrt{V}}\sum_{\vec k}\left(
\left[ \begin{array}{c} A_{\sigma +}(k\uparrow;t)\\ A_{\sigma -}(k\uparrow;t) \end{array} \right]\otimes w(k\uparrow)+
\left[ \begin{array}{c} A_{\sigma -}(k\downarrow;t)\\ A_{\sigma +}(k\downarrow;t) \end{array} \right]\otimes w(k\downarrow)\right. \nonumber\\
\left. +\left[ \begin{array}{c} B_{\sigma -}^{\dagger}(-k\uparrow;t)\\ B_{\sigma +}^{\dagger}(-k\uparrow;t) \end{array} \right]\otimes w(k\uparrow)
+\left[ \begin{array}{c} B_{\sigma +}^{\dagger}(-k\downarrow;t) \\ B_{\sigma -}^{\dagger}(-k\downarrow;t)  \end{array} \right]\otimes w(k\downarrow)\right)e^{i\vec k\vec x}.
\label{b.6}\end{eqnarray}
Under the conditions among $\{\alpha_{j},\beta^{\dagger}_{j}\}$ and their Hermitian conjugates
\begin{eqnarray}
\{\alpha_{j}(kr;t),\alpha^{\dagger}_{l}(k's;t)\}&=&\{\beta_{j}(kr;t),\beta^{\dagger}_{l}(k's;t)\}=\delta_{jl}\delta_{rs}\delta(\vec k-\vec k'),\nonumber\\
\mbox{others}&=&0,
\label{b.7}\end{eqnarray}
we obtain
\begin{eqnarray}
\{A_{\sigma\pm}(kr;t),A^{\dagger}_{\rho\pm}(k's;t) \}&=&\{B_{\sigma\pm}(kr;t),B^{\dagger}_{\rho\pm}(k's;t) \}=\sum_{j}U_{\sigma j}U_{\rho j}^{*}\frac{E_{j}\pm k}{2E_{j}}\delta_{rs}\delta(\vec k,\vec k'),\nonumber\\
\{A_{\sigma\pm}(kr;t),A^{\dagger}_{\rho\mp}(k's;t) \}&=&\{B_{\sigma\pm}(kr;t),B^{\dagger}_{\rho\mp}(k's;t) \}=\sum_{j}U_{\sigma j}U_{\rho j}^{*}\frac{m_{j}}{2E_{j}}\delta_{rs}\delta(\vec k,\vec k'),\nonumber\\
\mbox{others }&=&0.
\label{b.8}\end{eqnarray}
In the extremely relativistic limit, we have
\begin{eqnarray}
\{A_{\sigma +}(kr;t),A^{\dagger}_{\rho +}(k's;t) \}&\longrightarrow &\delta_{rs}\delta(\vec k,\vec k'),\nonumber\\
\{B_{\sigma +}(kr;t),B^{\dagger}_{\rho +}(k's;t) \}&\longrightarrow &\delta_{rs}\delta(\vec k,\vec k'),\label{b.9}\\
\mbox{\hspace*{5mm} others \hspace*{5mm}}&\longrightarrow &0;\nonumber
\end{eqnarray}
the high momentum part in r.h.s. of (\ref{b.6})
\begin{eqnarray}
\sim \frac{1}{\sqrt{V}}\sum_{\vec k}\left(
\left[ \begin{array}{c} A_{\sigma +}(k\uparrow;t)w(k\uparrow)\\ A_{\sigma +}(k\downarrow;t)w(k\downarrow) \end{array} \right]+\left[ \begin{array}{c} B_{\sigma +}^{\dagger}(-k\downarrow;t)w(k\downarrow)\\ B_{\sigma +}^{\dagger}(-k\uparrow;t)w(k\uparrow) \end{array} \right]\right)e^{i\vec k\vec x}
\label{b.10}\end{eqnarray}
with
\begin{eqnarray}
A_{\sigma +}(kr;t)=\sum_{j}U_{\alpha j}\alpha_{j}(kr;t), \mbox{\hspace*{5mm}}
B^{\dagger}_{\sigma +}(-kr;t)&=&\sum_{j}U_{\alpha j}\beta^{\dagger}_{j}(-kr;t).
\label{b.11}\end{eqnarray}
According to the assertion in ref.\cite{giunti}, one may construct an approximate Fock space of weak states in terms of $\{A^{\dagger}_{\sigma +}(kr;t),B^{\dagger}_{\sigma +}(-kr;t) \}$ give by (\ref{b.11}) only in the case of extremely relativistic neutrinos.
This assertion is in some sense self-evident, since masses of neutrinos become irrelevant in the extremely relativistic neutrinos.

It is more natural, however, to define, instead of (\ref{b.11}), a set of independent operators
\begin{eqnarray}
\tilde A_{\sigma}(kr;t)&:=&\sum_{j}U_{\sigma j}\left[\alpha_{j}(kr;t)\kappa_{j+}-i\beta^{\dagger}_{j}(-kr;t)\kappa_{j-}\right]\nonumber\\
\tilde B^{\dagger}_{\sigma}(-kr;t)&:=&\sum_{j}U_{\sigma j}\left[-i\alpha_{j}(kr;t)\kappa_{j-}+\beta^{\dagger}_{j}(-kr;t)\kappa_{j+}\right]
\label{b.12}\end{eqnarray}
so that the independent number of $\{\tilde A_{\sigma},\tilde B^{\dagger}_{\sigma}\}$ are equal to that of $\{\alpha_{j},\beta^{\dagger}_{j}\}$.
We obtain
\begin{eqnarray}
\{\tilde A_{\sigma}(kr;t),\tilde A^{\dagger}_{\rho}(k's;t)\}&=&\delta_{rs}\delta(\vec k,\vec k')=\{\tilde B_{\sigma}(-kr;t),\tilde B^{\dagger}_{\rho}(k's;t)\},\nonumber\\
\mbox{\hspace*{5mm} others \hspace*{5mm}}&=&0.
\label{b.13}\end{eqnarray}
In ref.\cite{giunti}, a set of operators, the number of which is twice times larger than the independent number, are introduced (similar to (\ref{b.5}) in the case of Dirac neutrinos).
Instead of doing so, we have defined a set of necessary and sufficient number of independent operators (\ref{b.12}) as a kind of Bogolyubov transformation, and obtained the canonical commutation relations (\ref{b.13}), irrespective of the value of $|\vec k|$.

From (\ref{b.4}) we obtain
\begin{eqnarray}
\nu_{\sigma}(x)=\frac{1}{\sqrt{V}}\sum_{\vec k}\left[\left(\tilde A_{\sigma}(k\uparrow;t)\left[ \begin{array}{c} 1\\0  \end{array} \right]+i\tilde B_{\sigma}^{\dagger}(-k\uparrow;t)\left[ \begin{array}{c} 0\\1  \end{array} \right]\right)\otimes w(k\uparrow)\right.\nonumber\\
\left.+\left(\tilde A_{\sigma}(k\downarrow;t)\left[ \begin{array}{c} 0\\1  \end{array} \right]+i\tilde B_{\sigma}^{\dagger}(-k\downarrow;t)\left[ \begin{array}{c} 1\\0  \end{array} \right]\right)\otimes w(k\downarrow)\right]e^{i\vec k\vec x}.
\label{b.14}\end{eqnarray}
The important feature of this expansion is that the operators $\{\tilde A_{\sigma},\tilde B^{\dagger}_{\sigma}\}$ depend explicitly on $m_{j}$'s as well as $|\vec k|$, but not on $m_{\sigma}$'s explicitly.
In the case of 2-flavors(i.e. $\sigma=e \mbox{ and }\mu$), we have
\begin{eqnarray}
\left( \begin{array}{c} \tilde A_{e}(kr;t)\\ \tilde A_{\mu}(kr;t) \\ \tilde B^{\dagger}_{e}(-kr;t) \\ \tilde B^{\dagger}_{\mu}(-kr;t)\end{array} \right)=\tilde{\cal G}(\theta,k)\left( \begin{array}{c} \alpha_{1}(kr;t)\\ \alpha_{2}(kr;t) \\ \beta^{\dagger}_{1}(-kr;t) \\ \beta^{\dagger}_{2}(-kr;t)\end{array} \right)
\label{b.15}\end{eqnarray}
with
\begin{eqnarray}
\tilde{\cal G}(\theta,k)=\left( \begin{array}{cccc}
c_{\theta}c_{1}&s_{\theta}c_{2}&-ic_{\theta}s_{1}&-is_{\theta}s_{2}\\
-s_{\theta}c_{1}&c_{\theta}c_{2}&is_{\theta}s_{1}&-ic_{\theta}s_{2}\\
-ic_{\theta}s_{1}&-is_{\theta}s_{2}&c_{\theta}c_{1}&s_{\theta}c_{2}\\
is_{\theta}s_{1}&-ic_{\theta}s_{2}&-s_{\theta}c_{1}&c_{\theta}c_{2}
\end{array} \right),\mbox{\hspace*{5mm}}
c_{j}=\kappa_{j+}, s_{j}=\kappa_{j-}.\label{b.16}
\end{eqnarray}
(\ref{b.16}) is seen to be equal to ${\cal G}(\theta,k)$, (\ref{a.7}), in which the mass parameters $m_{e}$ and $m_{\mu}$ employed in the plane-wave expansion of the flavor neutrino fields $\nu_{\sigma}$'s are set equal to zeros. 
Because, as seen from (\ref{b.3}), we have e.g. $u(k\uparrow)$ for the zero mass is
\begin{eqnarray}
u(k\uparrow)=\left( \begin{array}{c} 1\\ 0 \end{array} \right)\otimes w(k\uparrow).
\label{b.17}\end{eqnarray}


\end{document}